# Optical trapping, manipulation, and 3D imaging of disclinations in liquid crystals and measurement of their line tension

Ivan I. Smalyukh[*], Bohdan I. Senyuk, Sergij V. Shiyanovskii, and Oleg D. Lavrentovich

The Liquid Crystal Institute and Chemical Physics Interdisciplinary Program, Kent State University, Kent, OH 44242-0001

Andrey N. Kuzmin, Alexander V. Kachynski, and Paras N. Prasad

The Institute for Lasers, Photonics, and Biophotonics, University at Buffalo, The State University of New York, Buffalo, NY 14260-3000

We demonstrate optical trapping and manipulation of defects and transparent microspheres in nematic liquid crystals (LCs). The three-dimensional director fields and positions of the particles are visualized using the Fluorescence Confocal Polarizing Microscopy. We show that the disclinations of both half-integer and integer strengths can be manipulated by either using optically trapped colloidal particles or directly by tightly-focused laser beams. We employ this effect to measure the line tensions of disclinations; the measured line tension is in a good agreement with theoretical predictions. The laser trapping of colloidal particles and defects opens new possibilities for the fundamental studies of LCs.

[*] Corresponding author. Tel.: +1-330-672-1518. Fax.: 1-330-672-2796. E-mail: Smalyukh@lci.kent.edu

## 1. INTRODUCTION

The study of topological defects is important for understanding defect-mediated phase transitions in condensed matter [1-5]; it is also of a great current interest in cosmology [6-9]. The nematic liquid crystals (LCs) are convenient model systems for the study of defects and can be useful even for the "cosmology in laboratory" experiments [6-9]. Long-range orientational order in nematic LCs is usually described by the director field $\hat{n}(\vec{r}) = -\hat{n}(\vec{r})$ representing the average spatial orientation of the LC molecules [1]. This order can be locally broken (so that the director field $\hat{n}(\vec{r})$ can not be defined) on a point, along a line, or along a wall, giving rise to defects. The line defects in nematics are called disclinations; they are classified according to their strength $m$ that is defined as a number of revolutions by $2\pi$ that the director $\hat{n}(\vec{r})$ makes around the defect core when one circumnavigates the core once [1]. While the half-integer defect lines are topologically stable, the integer-strength disclinations are not and usually relax into nonsingular configurations of $\hat{n}(\vec{r})$. Disclinations can be obtained by a rapid quench from high-symmetry isotropic to the lower-symmetry nematic phase; it is much more difficult to spatially control (manipulate) them, a capability that would be desired for the topological study and "cosmology-in-laboratory" experiments [1-9]. The field-induced propagation [2] and deformation [3] dynamics of disclinations allow one for a limited spatial manipulation of these defects but only at certain boundary conditions; more robust approaches to manipulate defects are in a great demand.

Since the pioneering experiments of Ashkin [10], the laser tweezers were shown to be capable of trapping and manipulation of different transparent objects (particles) in the bulk of different isotropic media. Optical trapping became a valuable tool in physics and biology [11-15]. Laser tweezers allow one to measure pico-Newton forces associated with colloidal interactions [11] and unfolding of biopolymer molecules (by pulling a bead attached to one of the ends of the molecule) [12]. Optical trapping is often described in terms of competition of the scattering force, $F_{Scat}$, arising from backscattering of photons and pushing the object along the optical axis, and the gradient force, $F_{Grad}$, which pulls the object in the direction of increasing intensity of the beam, Fig.1a [14, 15], provided that the refractive index of the particle is larger than the refractive index of the surrounding medium. In a laser beam tightly focused by a high numerical aperture objective, there are substantial intensity gradients both perpendicular to the beam axis and along this axis which are pulling the bead towards the focus with the highest laser intensity, Fig.1a. When $F_{Grad}$ overcomes $F_{Scat}$, a stable trapping of a bead is possible. In the case of trapping with the Gaussian laser beams, a bead must have a higher refractive index than that of the surrounding medium in order to satisfy the $F_{Grad} > F_{Scat}$ condition of stable trapping [14, 15]. Most of the trapping and manipulation experiments using laser tweezers were done for the isotropic fluids. The advantages of laser tweezers in the study of LCs only begin to be explored [16-22] and the important new features of laser trapping in these birefringent media begin to be realized [17,19]; for example, optical trapping allowed researches to study both dipolar [17,18] and quadrupolar [23] colloidal interactions of particles immersed in nematic LCs.

We describe the optical trapping and manipulation of disclinations in nematic LCs. We show that the defect lines of both half-integer and integer strengths can be manipulated by either using optically trapped colloidal particles or directly by tightly-focused polarized laser beams. The Fluorescence Confocal Polarizing Microscopy (FCPM) [24] allows us to visualize the three-dimensional director fields and spatial positions of the particles and defects. We employ the manipulation with the optically trapped colloidal particles to measure line tension properties of the nematic disclinations; the experimental data are in a good agreement with theoretical predictions [1].

## 2. EXPERIMENT

### 2.1. Materials and cell preparation

We used the nematic LC mixtures ZLI2806 with small birefringence $\Delta n \approx 0.04$ and average refractive index $\bar{n}_{LC} = \sqrt{(2n_o^2 + n_e^2)/3} \approx 1.49$ ($n_o = 1.48$ and $n_e = 1.52$ are ordinary and extraordinary refractive indices, respectively) and E7 with a comparably large birefringence $\Delta n \approx 0.23$, $\bar{n}_{LC} \approx 1.6$ (both materials were obtained from EM Chemicals). The LCs were doped with 0.01wt.% of fluorescent dye n,n'-bis(2,5-di-tert-butylphenyl)-3,4,9,10-perylenedicarboximide (BTBP) for the FCPM studies [24]. We used Melamine Resin (MR) spherical particles (purchased from Aldrich) of refractive index $n_{MR} \approx 1.68$ and diameters $D = 3\mu m$ and $D = 4\mu m$ and polystyrene (PS) microspheres (obtained from Duke Scientific) with $n_{PS} \approx 1.6$ and $D = 1\mu m$. The MR microspheres were tagged with the Rhodamine B dye (maximum of single-photon absorption at ~540 nm) which fluoresced when excited by the $\lambda$=568 nm Kr-laser in the FCPM experiment, and also when trapped by $\lambda$=1064 nm laser tweezers (the later due to two-photon absorption). Both LCs and polymer particles are transparent at the wavelength of optical trapping ($\lambda$=1064nm), as needed to avoid heating of the samples.

The cells were constructed from glass plates of small thickness (150$\mu m$) and sealed using UV-curable glue; the cell gap was set either by Mylar spacers at the cell edges or using comparably big (tens of microns) glass particles dispersed within the cell area. The cell gap was varied within the range 20-100 $\mu m$. Surfaces of some of the substrates were coated with thin polyimide layers to set the direction of the easy axes at the LC-glass interfaces. The LC containing small concentration (<0.02% by weight) of well-separated polymer particles (we used sonication) was introduced into the cells by capillary forces.

We obtain disclinations of different strengths by cooling the cells from the isotropic phase. The defect lines of opposite strengths quickly annihilate with time soon after the LC is cooled below the isotropic-nematic transition. Some of the defects are trapped on spacers and irregularities or stabilized by boundary conditions at the confining surfaces; these disclinations will be a subject of our study.

## 2.2. Fluorescence Confocal Polarizing Microscopy

Using the two-channel FCPM [24], we determine the director configurations in the cell (through the fluorescence of BTBP detected in the spectral range 510-550nm, channel #1) as well as the positions of the MR particles (detecting fluorescence from Rhodamine B dye embedded in the spheres in the spectral range 585-650nm, channel #2). We used Olympus Fluoview BX-50 confocal microscope, modified by a linear polarizer $\hat{\mathbf{P}}$ that sets polarization $\hat{\mathbf{P}}_e \parallel \hat{\mathbf{P}}$ of the excitation beam and the polarization $\hat{\mathbf{P}}_f \parallel \hat{\mathbf{P}}$ of the detected fluorescent light. The excitation beams of a 488 nm Ar laser and a 568 Kr laser are focused by a high NA objective into a small ($<1\,\mu m^3$) volume in the LC cell. The fluorescent light from this volume is detected by two photomultiplier tubes that collect light in the spectral ranges 510-550nm for channel #1 and 585-650nm for channel #2 as selected by the interference filters. A 100 μm wide pinhole discriminates against the regions above and below the selected volume, which allows for the diffraction-limited resolution along the optical axis of the microscope. The strong orientation dependence of measured fluorescence signal allows us to decipher the 3D orientation pattern from the FCPM observations. The resolution of the FCPM is ~1μm in both radial (in plane of the LC cell) and axial (along the cell normal $\hat{Z}$) directions.

## 2.3. Laser tweezers

We use a dual beam laser trapping system consisting of an optical manipulator (Solar-TII, LM-2), a TEM$_{00}$ CW Nd: YAG laser (COHERENT, Compass 1064-2000) with $\lambda=1064$ nm, and a modified inverted microscope (NIKON, TE-200) [13, 17]. An optical trap is formed by a 100× microscope objective (NA=1.3); this trap is used for 3D positioning of particles within the cell. The submicron waist size of the focused light beam ($\approx 0.8\mu m$) is smaller than diameters of the used polymer microspheres, which allows us to avoid reorientation of LC molecules under the intense laser radiation and its influence on the results of the line tension measurements. Before each measurement, we perform force calibration in the studied sample and at the particular depth of trapping. The focused beam is steered in the horizontal plane (plane of the cell) by a computer-controlled galvano-mirror pair; the vertical coordinate of the trap is controlled by a piezo-stage with the accuracy 0.1 μm.

## 3. RESULTS

### 3.1. Optical trapping and manipulation of particles in liquid crystals

Laser tweezers experiments in anisotropic fluids (such as LCs) require some additional care because (a) the refractive index difference between the particle and the host medium depends on the local LC director field $\hat{n}(\vec{r})$ and light polarization, (b) birefringence of LC results in light defocusing, and (c) the focused light beam can reorient the local $\hat{n}(\vec{r})$ as demonstrated for the nematic LC in flat slabs [20] and spherical droplets [22]. These problems are intrinsic when applying laser tweezers for the study of anisotropic media; they can be mitigated by using LCs with small birefringence $\Delta n$ and colloidal particles larger than the waist of the laser beam, Fig.1b, as we will show below.

The polymer particles in LCs can be stably trapped and manipulated by steering the laser beams, Fig.2. Moreover, one can assemble different kinds of colloidal structures. Fig. 2d is a qualitative illustration of optical manipulation of colloidal particles in the nematic bulk: the letters "LC" have been assembled by tweezing the $3\mu m$ MR beads suspended in the LC host (ZLI2806).

In order to characterize the trapping forces acting on the colloidal particles in the nematic bulk, we use the technique based on viscous drag forces exerted by the LC [17]. A laser-trapping beam forces the particle to move with an increasing linear velocity $V$ until the viscous force overcomes the trapping force and the bead escapes the optical trap. During the calibration, the beads are steered along linear paths in planar cells (the escape velocity $V_e$ was found to be about two times larger when moving the bead along the director, $\vec{V} \| \hat{n}(\vec{r})$, than when moving it perpendicular to the director, $\vec{V} \perp \hat{n}(\vec{r})$) or along circular paths in the homeotropic cells and when the sample is heated to the isotropic phase. The trapping force $F_t$ is calculated using the Stoke's law $F_t = 3\pi D \alpha_4 V_e$, where $V_e$ is average escape velocity, and the effective viscosity of the LC is approximated by the Leslie coefficient $\alpha_4$ [1, 25]. The Reynolds numbers are verified to be low enough ($\leq 10^{-5}$) to justify the use of the Stokes law [1, 13, 14]. We determine the average escape velocity $V_e$ and calibrate the trapping force $F_t$ for different laser powers, Fig.3; this calibration we will use for the measurement of defect line tensions in the section 3.3 below.

Similarly to the case of isotropic fluids, the optical trapping forces in LCs increase with the laser power, Fig. 3(a). These forces in ZLI2806 are weaker than for the same particles dispersed in water (which is natural as the difference between the refractive index of the particle and the surrounding medium is larger in the case of water) but still sufficiently strong to enable laser manipulation. The spherical aberrations that often arise due to the refractive index mismatch at the coverslip-sample interfaces can considerably weaken the trapping forces, especially if one uses the oil immersion objectives for trapping in the low-refractive-index fluids such as water [15]. Because of the spherical aberrations, the spatial size of the focused light spot increases with the depth of trapping and the originally sharp intensity distribution is blurred. The spherical aberration effect on the trapping in the LC is small as the average refractive index of LCs is close to that of a silica glass, $\sim 1.5$. The efficiency of trapping in ZLI2806 with a small birefringence $\Delta n$ does not decrease much when the depth of scanning increases, contrary to the case of water, Fig.3(b). The particles suspended in ZLI2806 could be trapped even at a depth of $80 \mu m$. However, if $\Delta n$ is significant, as in the case of E7 with high birefringence $\Delta n \approx 0.23$, then substantial light defocusing weakens the trapping forces, Fig.3(b). When the trapping depth reaches about $20 \mu m$, optical manipulation in E7 becomes practically impossible, Fig.3(b). This result is natural, as in the birefringent media light defocusing (and the spatial dimensions of the optical trap) increases with the depth of focusing and with birefringence $|\Delta n|$ of the LC [24].

The trapping forces can also depend on the director field along the path of light if the beads are manipulated in the E7 cells; practically no such dependence is observed if the host medium is ZLI2806 with small birefringence. In general, high $\Delta n$ makes quantitative measurement of colloidal interaction in the LC such as E7 very difficult even though trapping and manipulation of beads in these media is often still possible (if the depth of trapping is not too large). However, low $\Delta n$ materials such as ZLI2806 are perfectly suited for optical trapping and can be even used to study colloidal interactions and parameters of topological defects in the nematic phase, as we demonstrate below.

### *3.2. 3D imaging, trapping, and manipulation of defects in liquid crystals*

Nematic disclinations of opposite charges have similar appearance under a microscope and, especially if they are parallel to the plane of observations, it is often difficult to distinguish the $m=-1/2$ from $m=1/2$ as well as the $m=-1$ from $m=1$ defect lines. We use the FCPM technique to map the 3D director around the disclination which allows us to determine the strengths of defect lines. Imaging of the director field associated with disclination is illustrated on the example of a straight disclination with its ends pinned at two spacers that were used to set the cell thickness, Fig.4. FCPM vertical cross-section obtained with linearly polarized probing light, Fig.4(a), reveals the director reorientation by $\pi$ around the disclination as one circumnavigates the core once, Fig.4(a). This director configuration is assisted by the type of anchoring in the cell, which was tangential degenerate at the top plate and homeotropic at the bottom plate. As it is revealed by the two-channel FCPM imaging, both the disclination and the colloidal particle that can be used to manipulate the defect line are located in the nematic bulk, Fig.4(a-c). The co-localization of the fluorescent signals from the dye-doped LC (Fig.4a, represented using a color-coded intensity scale) and the Rhodamine B stained MR particle (Fig.4b) shows that the particle is located in the close vicinity of the defect core, Fig.4c. The defect lines studied below, Figs. 5-8, look similar under the polarizing microscope; the visible difference which is usually used to distinguish them is that the half-integer disclinations, Figs. 7,8 are thinner than the integer-strength disclinations, Figs. 5,6 (compare the thickness of the defect lines to the diameter of the 3 $\mu m$ beads). Using the FCPM technique and mapping $\hat{n}(\vec{r})$ around a defect as described above, we determined not only the absolute value of the disclination strength m but also its sign (the strengths of disclinations are marked in Figs. 5-8).

The strong director distortions and the associated gradients of the effective refractive index in the vicinity of the disclination cores allow for trapping and manipulation of these defects using laser tweezers. To illustrate this we use a straight horizontal *m*=1 disclination that joins two glass spacers in a flat cell, Fig.5. The disclination can be stretched by manipulating the laser beam focused at the disclination midway between the spacers, Fig.5. The local director $\hat{n}(\vec{r})$ in the center of the *m*=1 disclination is along the defect line, Fig.5d,e, making the defect core non-singular, i.e., escaped into the 3$^{rd}$ dimension. For a laser beam polarized along the disclination, the effective refractive index in the center of the defect line is $n_{eff,core} \approx n_e$, which is larger than the effective refractive index of the surrounding LC $\approx n_o$ (here $n_o = 1.48$ and $n_e = 1.52$ are ordinary and extraordinary refractive indices of the used ZLI2806, respectively). The refractive index difference between the core of the *m*=1 disclination and the LC around it is close to LC birefringence, $\Delta n \approx 0.04$, and is sufficient for manipulation of the disclination using laser tweezers, Fig. 5. As one can expect, the disclination is trapped when the laser beam is polarized along the defect line, Fig.5; the defect is repelled from the focused laser spot if the laser beam is polarized orthogonally to the disclination as in this case the effective refractive index difference between the defect core and surrounding LC becomes negative. In a similar way, the effective refractive index difference between the defect core and surrounding LC allows one to optically trap and manipulate other types of disclinations using tightly focused laser beams. We note that in addition to the mechanism of disclination trapping described above, the trapping and manipulation of disclinations might be also influenced by the LC realignment under the intense laser radiation. If laser power exceeds some critical value (about 30-50mW in the case of ZLI2806) and the direction of linear light polarization differs from the local director $\hat{n}(\vec{r})$, we notice spots of realigned LC at the place of focused infrared laser beam, similar to those observed in Ref. [19]. The effect of light-induced LC realignment on the manipulation of disclinations increases with laser power. Because of the nonlinear optical effects such as LC reorientation at laser irradiation, quantitative characterization of laser trapping of disclinations with high-power laser tweezers is difficult.

A disclination can be also manipulated by using an optically trapped polymer bead, Fig. 6. A particle with a diameter much larger than the anchoring extrapolation length of the particle-LC interface can be either attracted to or repelled from the defect core, depending on the type of disclination and surface anchoring conditions at the particle's surface. For example, pulling the MR spherical particle with the tangential anchoring conditions into the defect core of the $m=1$ disclination with a nonsingular defect core would require strong director transformation, Fig.6i; this is associated with an energetic barrier and explains why the optically trapped bead can not easily penetrate through the $m=1$ defect line. One might wonder why the disclination does not leave the particle and passes over the top/bottom of the bead. The most plausible reason is pinning at the surface. Similar effect of pinning of disclination ends has been demonstrated in the experiments of Ref. [5] in which the disclinations joined two plates that were rotated with respect to each other. Despite the fact that the alignment at the plates was tangentially degenerate, the ends moved with the plates (thus stretching the disclinations themselves) demonstrating the effect of pinning [5]. By manipulating a particle, the disclination with its ends pinned at glass spacers can be stretched similarly to the case of an elastic string, Fig.5. When laser light is switched off, the disclination straightens up and pushes the bead. This is natural as the energy of a straight disclination is proportional to its length and, in first approximation, also the energy of the curved defect line is proportional to its length [1]. Therefore, a curved defect line has a tendency to straighten, in order to decrease its length and total elastic energy. This tendency can be described in the terms of a line tension, defined as a ratio of the variation of elastic energy to the variation in length [1], which will be studied in Section 3.3.

An optically trapped MR particle also allows us to manipulate a half-integer $m=-1/2$ disclination, Fig. 7. The bulk half-integer disclination has a singular core of the size $r_c$ close to the molecular size (or somewhat larger), of the order of (1-10) nm for a typical nematic thermotropic material such as E7 and ZLI2806, see also [27]. The surface disclination can have a much wider core, of the order of the ratio of the bulk elastic constant $K$ and the surface anchoring coefficient $W$, $l = K/W$, which is normally in the range $0.1-1\mu m$ [1, 28]. The energy of the disclination at the surface of the particle can be lower than the energy of a disclination in the bulk, as discussed by R.B. Meyer [28]. The disclination located sufficiently close to the surface of a particle might transform into the surface disclination, i.e. into the line which core is located at the surface and is thus of the size $l \gg r_c$. In the end of this manipulation process the particle and $m=-1/2$ disclination separate; the defect line quickly straightens up and returns to its original location.

### 3.3. Measurements of the disclination line tension

Since the elastic constants in most nematics are $\sim 10 pN$, the optical trapping approach [17] is also well suited for the quantitative study of line tension of defects. This we demonstrate below on an example of a m=1/2 disclination. Neglecting the difference in the values of Frank elastic constants, the line tension of the topologically-stable half-integer disclination is approximated as [1]:

$$T_d = \frac{\pi}{4} K \ln \frac{L}{r_c} + T_c, \tag{1}$$

where $K$ is the average elastic constant, $L$ is the characteristic size of the system (sample thickness in our case), $r_c$ and $T_c$ are the radius and energy of the defect core. In the model of an isotropic (melted) core [1], $T_c \approx \pi K/4$ and $r_c \approx 10 nm$. With the average elastic constant for ZLI2806 $K \approx 12 pN$ and $L = 10-100\mu m$, Eq.(1) predicts $T_d = 65-85 pN$.

By moving the particle with the laser tweezers, Fig.8(a-c), the disclination can be pulled similarly to an elastic string. When the bead is released from the optical trap, the disclination straightens up in order to minimize its length and thus the total elastic energy, Eq.(1). Displacing the particle by some distance and gradually decreasing the laser power, we find the magnitude of the pulling force $F_p$ (equal to the optical trapping force $F_t$) for which $F_p = 2T_d \cdot \cos\theta_d$, where $\theta_d$ is the angle defined in Fig.8(d). For $\theta_d$ in the range $90-45°$, we find the line tension of disclination $T_d = 74 \pm 4 \, pN$, close to the estimate obtained from Eq.(1) and to the data obtained by observations of thermal fluctuations of the defect line [29].

## 4. CONCLUSIONS

We demonstrated that the laser tweezers can be used to trap and manipulate defects in the nematic phase of thermotropic liquid crystals. We measured the disclination line tension which is found to be in a good agreement with the theoretical predictions. Laser trapping and manipulation of defects opens new possibilities for the fundamental studies of liquid crystals as well as defects in general. In particular, optical manipulation and disclination line tension measurements might be very useful tools to study defects in biaxial nematic LCs [30-32], properties of which are much less understood as compared to the case of uniaxial nematics.

## 5. ACKNOWLEDGEMENTS

I.I.S., B.I.S, S.V.S, and O.D.L. acknowledge support of the NSF, Grant DMR-0315523. I.I.S. acknowledges the grant of the International Institute of Complex and Adaptive Matter (I2CAM) to support his participation in the 8$^{th}$ European Conference on Liquid Crystals in Sesto, Italy, and also the 2005 Fellowship of the Institute of Complex and Adaptive Matter. The research at Buffalo was supported by an AFOSR DURINT grant, number F496200110358. We thank Dr. L. Longa and Dr. Yu. Nastishin for discussions.

**FIGURE CAPTIONS**

**Fig. 1.** Schematic illustration of optical trapping of colloidal particles by tightly focused laser beams: waist of the beam can be (a) larger or (b) smaller than the diameter $D$ of the bead.

**Fig. 2.** Manipulation of colloidal particles in the nematic liquid crystal: (a,b,c) changing of the center-to-center separation between the beads by using two laser beams; (d) a structure of $D = 3\mu m$ particles in form of the letters "LC" assembled in the bulk of a liquid crystal cell using laser beams. The black bar on the inset indicates the rubbing direction at the plates of the planar nematic LC cell.

**Fig. 3.** Trapping force vs. laser power (a) and depth of trapping (b) in liquid crystals ZLI2806 and E7 as compared to water. A 100× immersion oil objective with NA=1.3 is used in the experiment.

**Fig. 4.** Vertical FCPM cross-sections with the fluorescent signal from (a) BTBP-doped LC and (b) Rhodamine B labeled particle (gray) and (c) schematic drawing of the director field with the bead in the vicinity of a disclination. The color-coded FCPM fluorescence intensity scale is shown in (d). The polarization of the FCPM probing light is marked by "P" and a white bar in (a). The BTBP dye in the LC was excited with a 488nm Ar-laser and the fluorescent light was detected in the spectral region 510-550nm; the Rhodamine B dye incorporated in MR microspheres was excited using the 568nm Kr-laser and the respective fluorescence signal was detected in the spectral range 585-650nm.

**Fig. 5**. Trapping and manipulation of a m=1 disclination that has a nonsingular core escaped into the 3$^{rd}$ dimension: (a,b,c) polarizing microscopy images of the disclination manipulated by the infrared (the laser spot is not visible on the images) laser beam and (d,e) schematic illustration of the director field in the vicinity of the disclination and its trapping by a focused laser beam.

**Fig. 6.** Manipulation of the m=1 disclination using an optically trapped polymer particle with tangential anchoring: (a-h) polarizing microscopy images of a disclination being stretched by a bead trapped with the laser tweezers; (i) a schematic illustration showing that a particle can not penetrate through the disclination core because of the tangential anchoring conditions on its surface that are not compatible with strong director distortions in the vicinity of the defect.

**Fig. 7.** Manipulation of the m=-1/2 disclination using a particle with tangential anchoring by pulling the bead in the direction perpendicular the defect line: (a,c-h) polarizing microscopy textures and (b) schematic drawing of the director field. The disclination that (a, b) was originally separated from the particle with tangential anchoring is (c) adsorbed at the particle surface and becomes a surface disclination with energy lower than that of the defect line in the bulk; (d-g) as the optically trapped bead is moved farther from the position of unstretched disclination, (h) the bead separates from the bulk m=-1/2 disclination that straightens up to minimize its total energy and returns to its original position.

**Fig. 8.** Manipulation of a disclination and measurement of its line tension: (a-c) stretching the $m = 1/2$ disclination in the nematic ZLI2806 with an optically trapped $D = 4 \mu m$ MR polymer bead; (d) force-line tensions balance that is used to measure the disclination line tension.

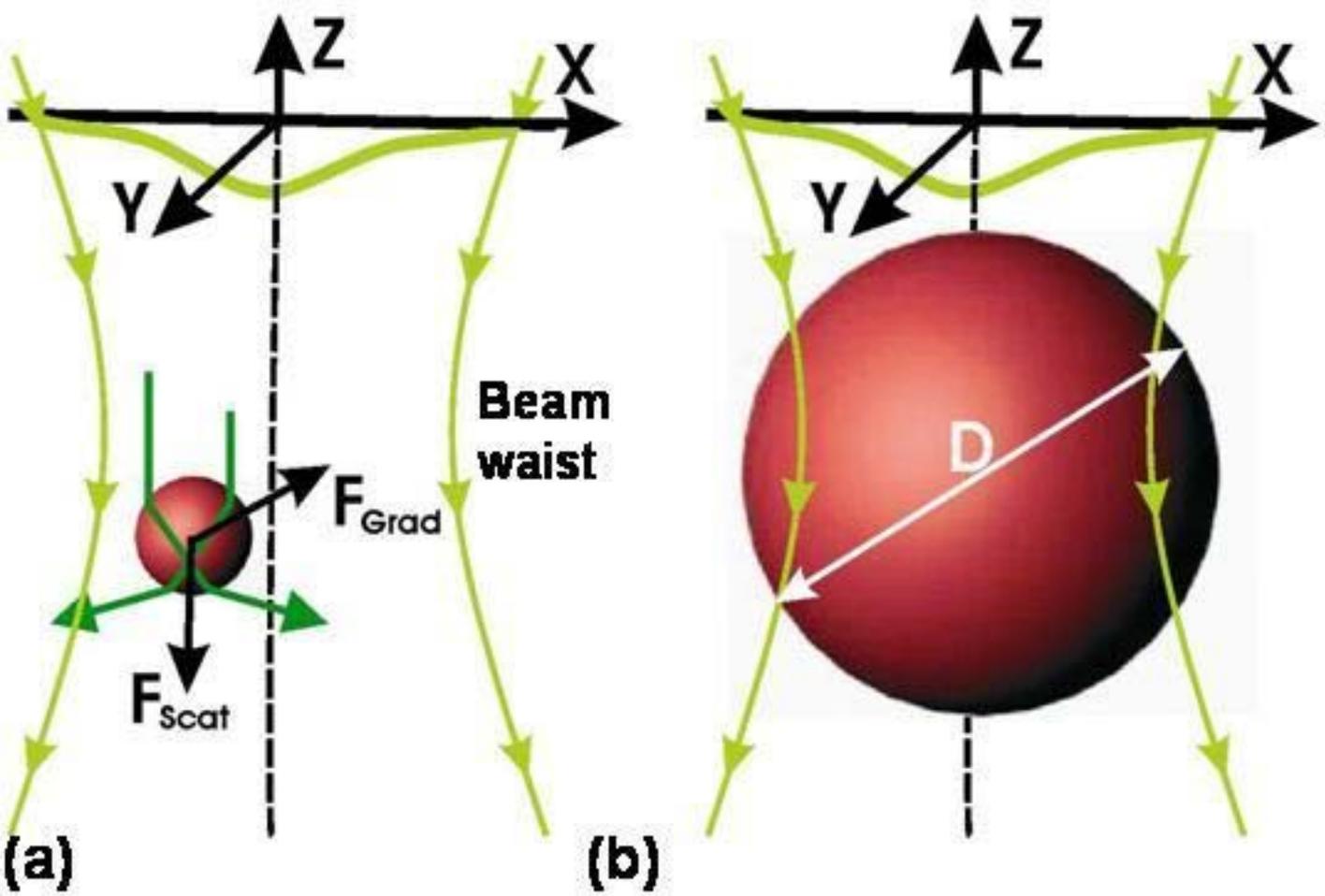

**Fig. 1.**

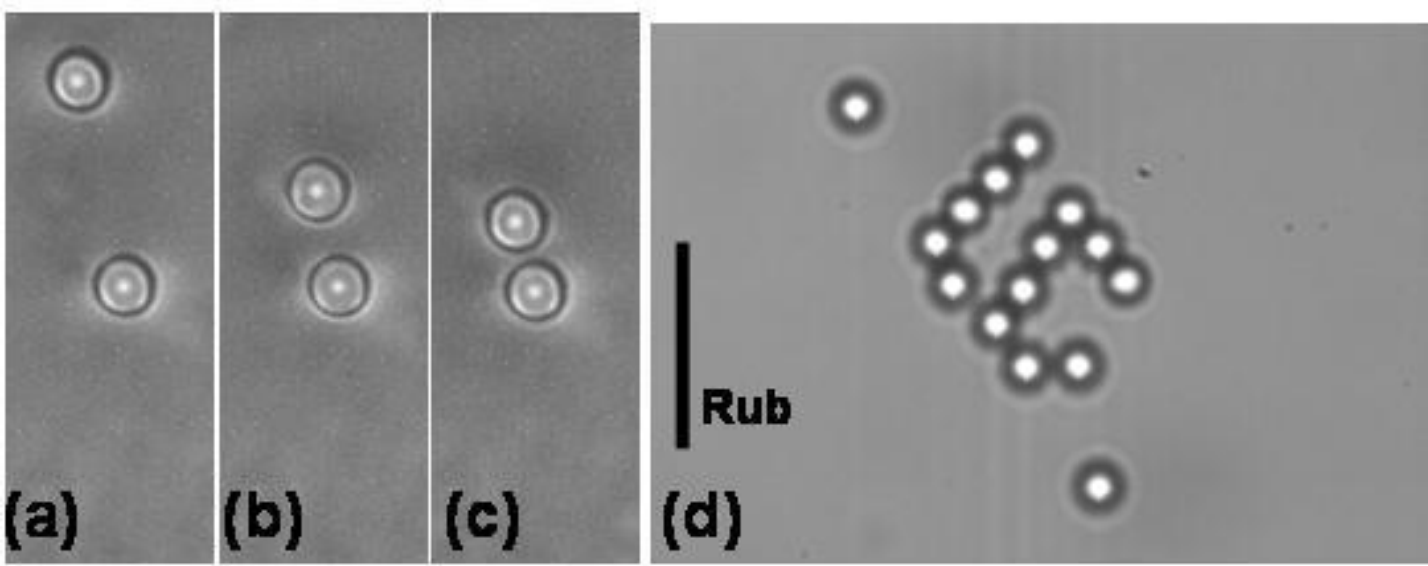

Fig. 2.

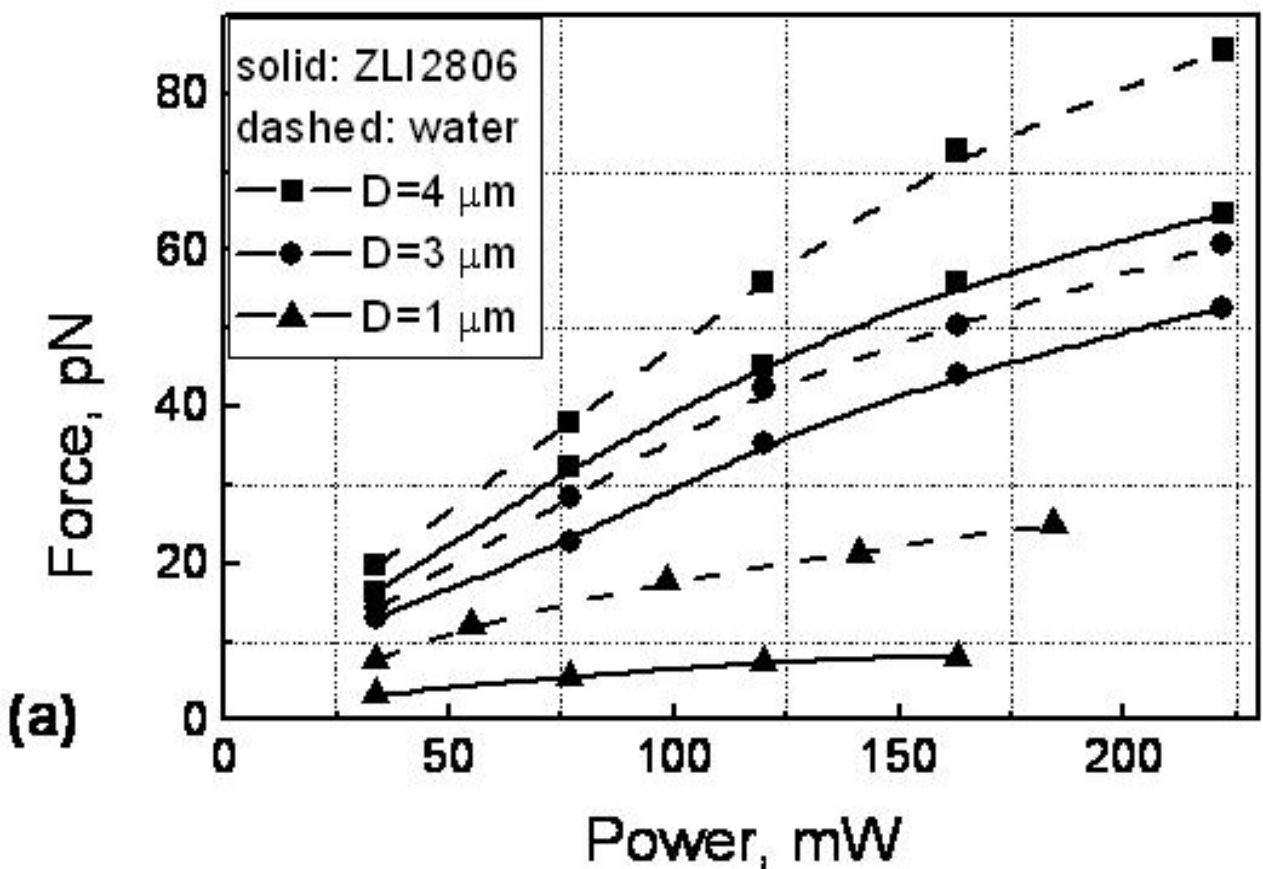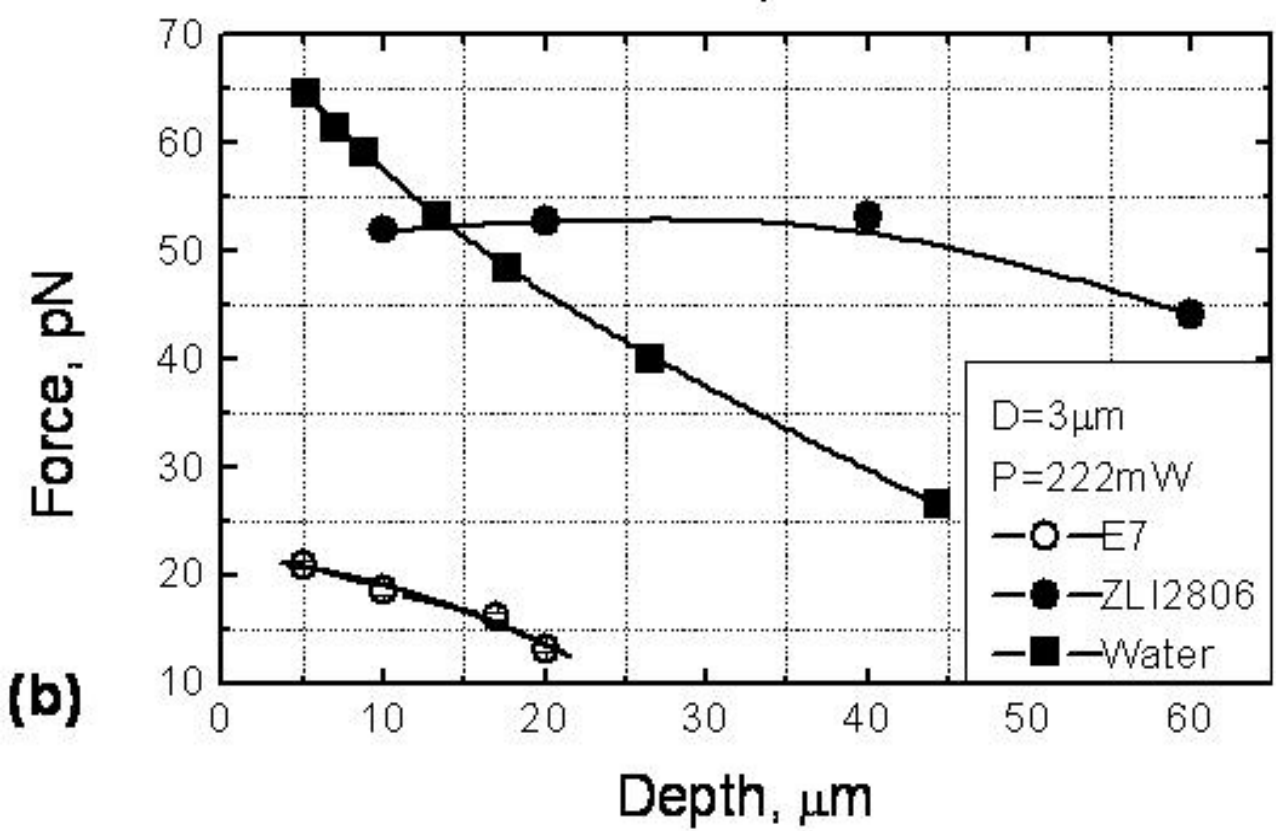

**Fig. 3.**

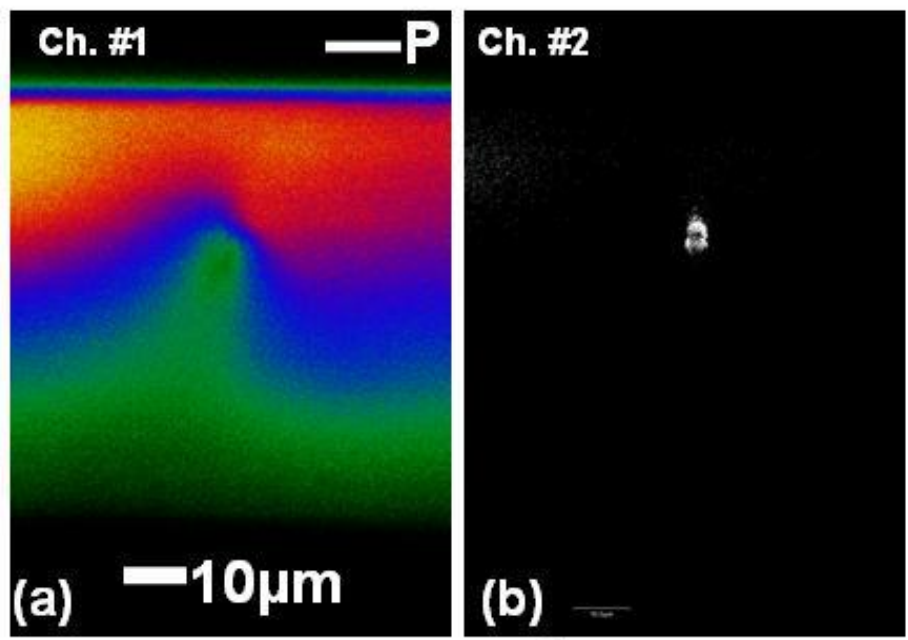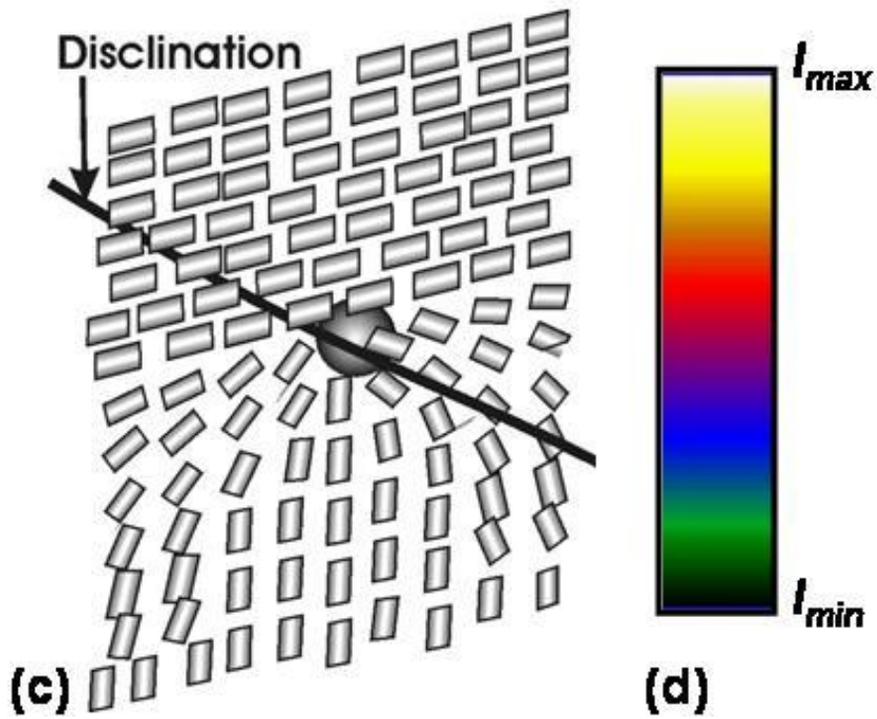

Fig. 4.

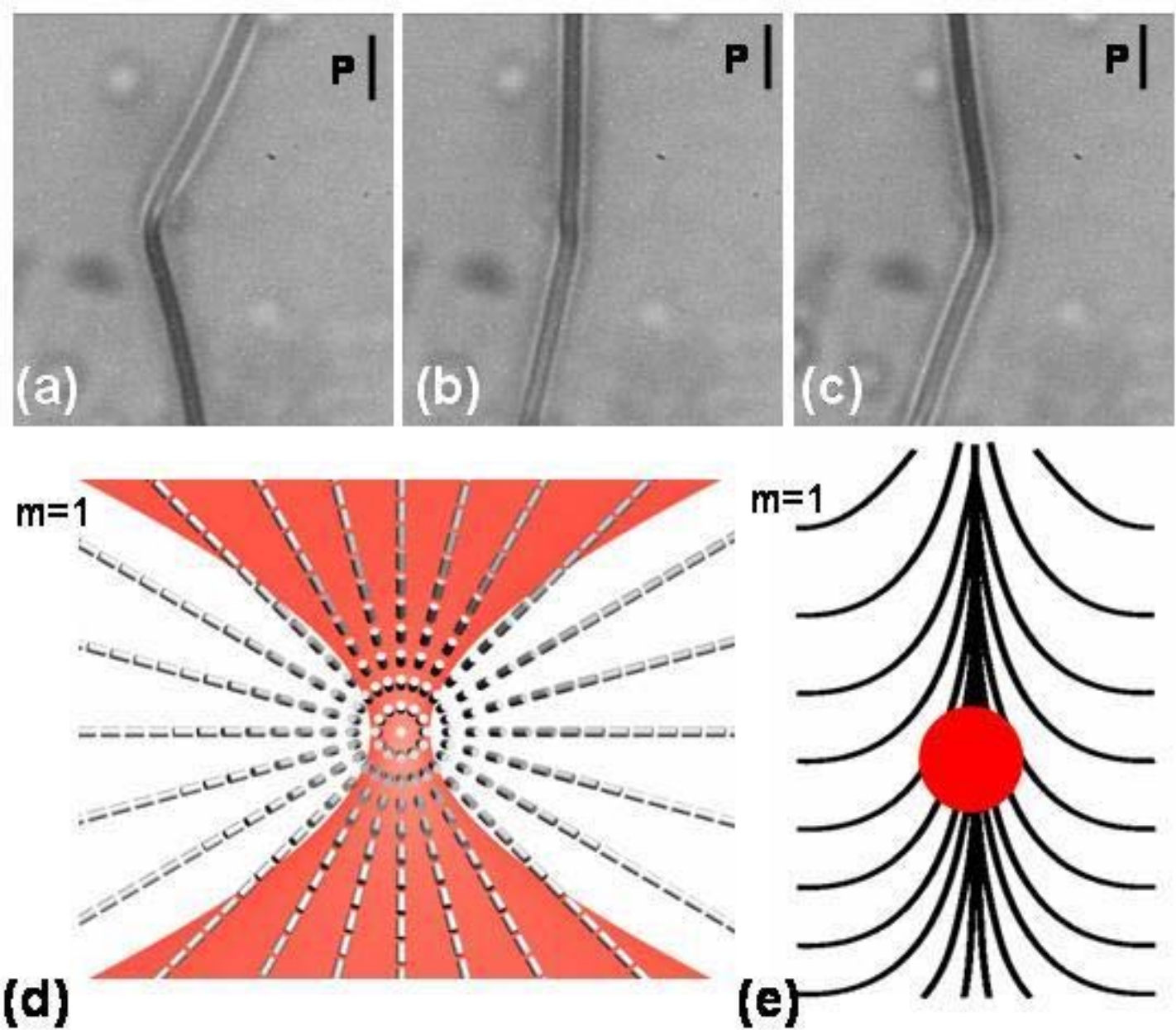

Fig. 5.

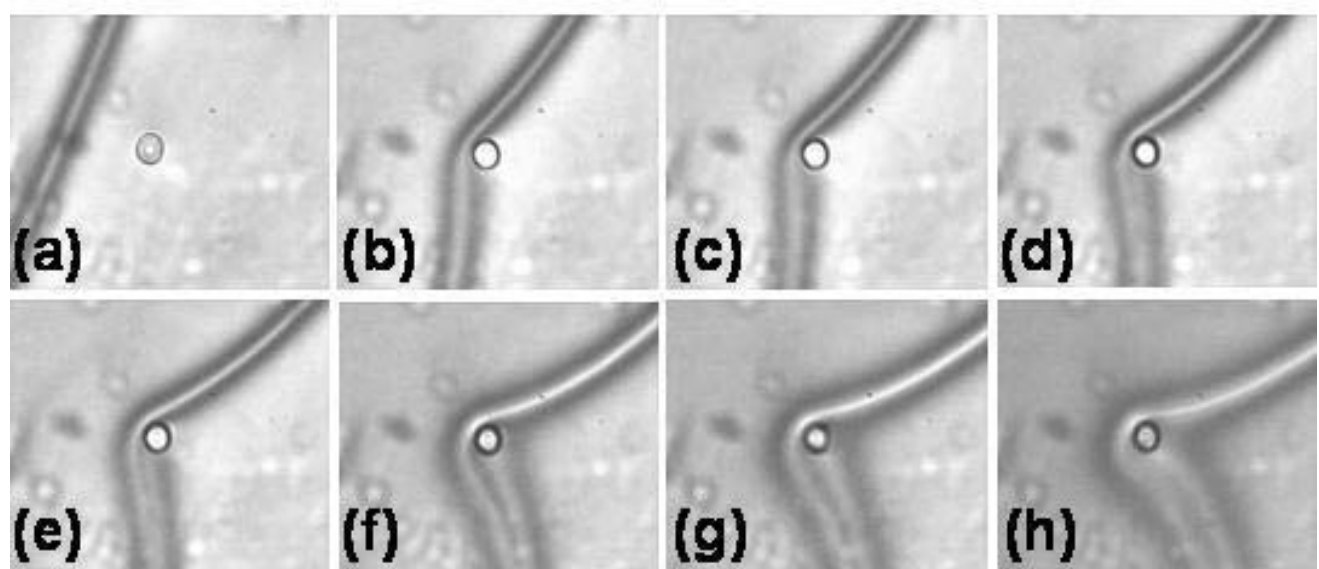
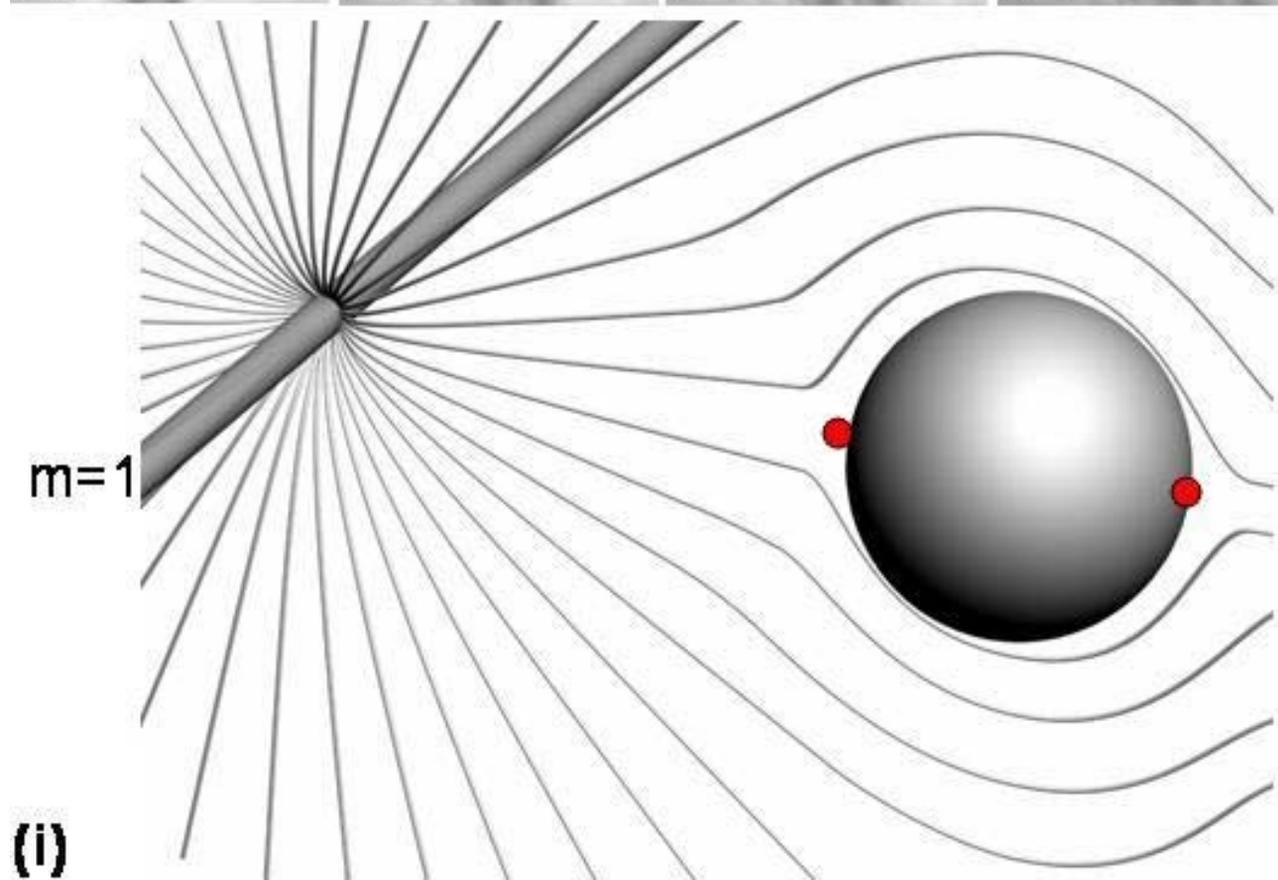

**Fig. 6.**

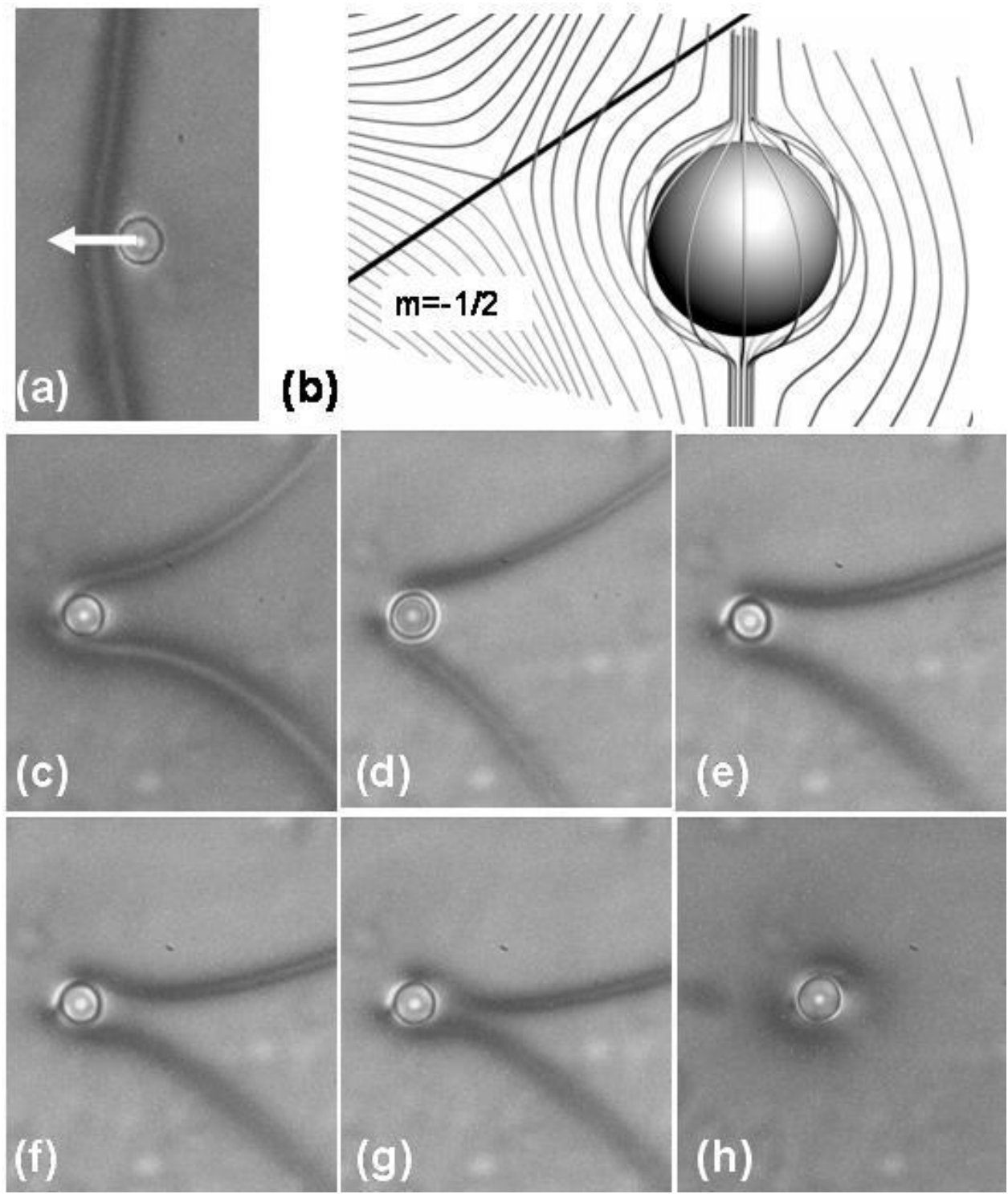

Fig. 7.

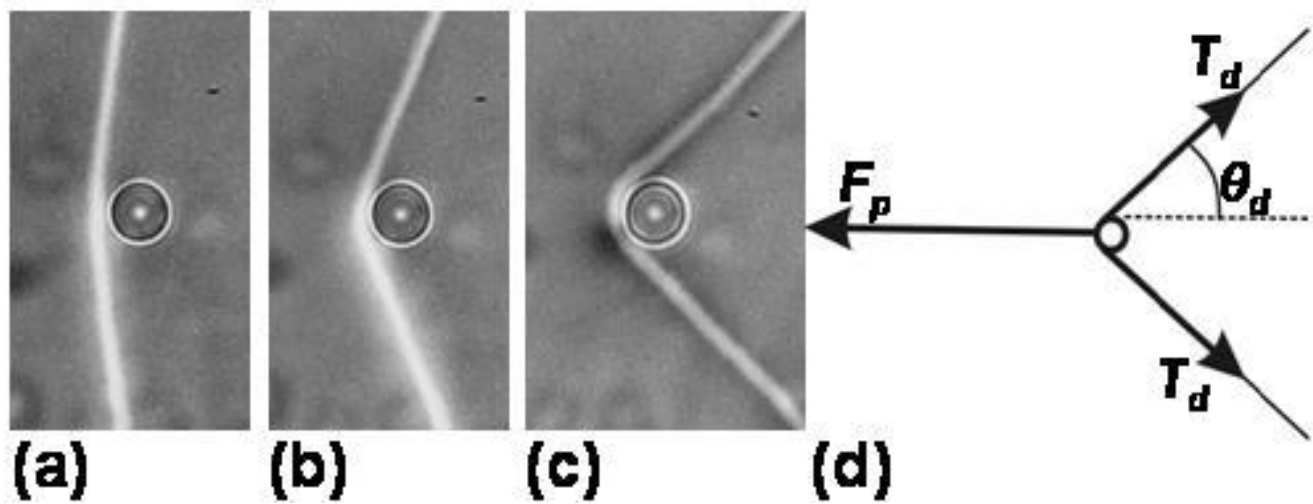

**Fig. 8.**